\documentclass[pra,a4paper,onecolumn,]{revtex4}
\usepackage[T1]{fontenc}
\usepackage{ae}
\usepackage{geometry}
\usepackage{amsmath}
\usepackage{graphicx}
\usepackage{subfigure}

\newcommand{\ket}[1]{\left|#1\right\rangle}
\newcommand{\psii}{\psi}
\newcommand{\bracket}[3]{\langle#1|#2|#3\rangle}
\newcommand{\kket}[1]{\ket{#1}}
\newcommand{\kotimes}{}

\begin{document}

\title{Experimental demonstration of a universally valid error-disturbance uncertainty relation in spin-measurements}
 \author{Jacqueline Erhart$^1$}
 \author{Stephan Sponar$^1$}
 \author{Georg Sulyok$^1$}
\author{Gerald Badurek$^1$}
\author{Masanao Ozawa$^2$}
\author{Yuji Hasegawa$^1$}
\email{Hasegawa@ati.ac.at}
 \affiliation{%
$^1$Atominstitut, Vienna University of Technology
Stadionallee 2, 1020 Vienna, Austria\\$^2$ Graduate School of Information Science, Nagoya University, Chikusa-ku, Nagoya 464-8601, JAPAN }

\date{\today}

\maketitle

\textbf{The uncertainty principle generally prohibits determination of certain pairs of quantum mechanical observables with arbitrary precision and forms the basis of indeterminacy in quantum mechanics\cite{Wheeler,Haroche}. It was Heisenberg who used the famous gamma-ray microscope thought experiment to illustrate this indeterminacy\cite{Heisenberg}. A lower bound was set for the product of the measurement error of an observable and the disturbance caused by the measurement. Later on, the uncertainty relation was reformulated in terms of standard deviations\cite{Kennard,Robertson}, which focuses solely on indeterminacy of predictions and neglects unavoidable recoil in measuring devices\cite{Ballentine}. A correct formulation of the error-disturbance relation, taking recoil into account, is essential for a deeper understanding of the uncertainty principle.
However, the validity of Heisenberg's original error-disturbance uncertainty relation is justified only under limited circumstances\cite{6,7,8}. Another error-disturbance relation, derived by rigorous and general theoretical treatments of quantum measurements, is supposed to be universally valid\cite{13,14}. Here, we report a neutron optical experiment that records the error of a spin-component measurement as well as the disturbance caused on another spin-component measurement. The results confirm that both error and disturbance completely obey the new, more general relation but violate the old one in a wide range of an experimental parameter.
}

The uncertainty relation was first proposed by Heisenberg\cite{Heisenberg} in 1927 as a limitation of simultaneous measurements of canonically conjugate variables due to the back action of the measurement: the measurement of the position $Q$ of the electron with the error $\epsilon(Q)$, or ``the mean error'', induces the disturbance $\eta(P)$, or ``the discontinuous change'',  of the momentum $P$ so that they always satisfy the relation
\begin{equation}
\epsilon(Q) \eta(P) \sim \frac{\hbar}{2}, \label{HEI-1}
\end{equation}
where $\hbar$ is Planck's constant divided by $2\pi$ (here, we use $\frac{\hbar}{2} $ for consistency with modern treatments). In a mathematical derivation of the above relation from the commutation relation $QP-PQ = i\hbar$, Heisenberg\cite{Heisenberg} used the reciprocal relation $\sigma(Q)\sigma(P)\geq\frac{\hbar}{2} $ for standard deviations $\sigma(Q)$, $\sigma(P)$ of position and momentum, which was proved shortly afterwards by Kennard\cite{Kennard} for arbitrary wave functions.
This relation was generalized to arbitrary pairs of observables $A$, $B$ by Robertson\cite{Robertson} as
\begin{equation}
    \sigma(A)\sigma(B)\geq\frac{1}{2}\left|\bracket{\psi}{[A,B]}{\psi}\right|
    \label{ROB}
\end{equation}
in any states $\kket{\psi}$ with $\sigma(A), \sigma(B) < \infty$. Here, $[A,B]$ represents the commutator $[A,B] = AB - BA$ and the standard deviation is defined as $\sigma(A)^2=\bracket{\psi}{A^2}{\psi} -\bracket{\psi}{A}{\psi}^2$.
Robertson's relation (Eq.~\ref{ROB}) for standard deviations has been
confirmed by many different experiments. In a single-slit diffraction experiment\cite{vor21} the uncertainty relation, as expressed in Eq.~\ref{ROB}, has been confirmed. The slit width determines the position spread while the diffraction pattern on the screen shows the momentum distribution: the pattern for a narrower slit gets wider and vise versa. A trade-off relation appears in squeezing coherent states of radiation fields\cite{22,23}. Starting with a theoretical proposal\cite{24} and the first experimental generation of squeezed states\cite{25}, many experimental demonstrations have been carried out\cite{26,27}.

Robertson's relation (Eq.~\ref{ROB}) has a mathematical basis, but has no immediate implications to limitations on measurements. This relation is naturally understood as limitations on state preparation or limitations on prediction from the past.
On the other hand, the proof of the reciprocal relation for the error $\epsilon(A)$ of an $A$-measurement and the disturbance $\eta(B)$ on observable $B$ caused by the measurement, in a general form of Heisenberg's error-disturbance relation
\begin{equation}
    \epsilon(A)\eta(B)\geq\frac{1}{2}|\bracket{\psi}{[A,B]}{\psi}|,
    \label{HEIS}
\end{equation}
is not straightforward, since Heisenberg's proof used
an unsupported assumption on the state just after the measurement\cite{27+1}. Recently, rigorous and general theoretical-treatments of quantum measurements have revealed the failure of Heisenberg's relation
(Eq.~\ref{HEI-1}),
and derived a new universally valid relation\cite{13,14} given by
\begin{equation}
    \epsilon(A)\eta(B)+\epsilon(A)\sigma(B)+\sigma(A)\eta(B)\geq\frac{1}{2}|
    \bracket{\psi}{[A,B]}{\psi}|.
    \label{OZA}
\end{equation}
\sloppy
Here, the error $\epsilon(A)$ is defined as the root-mean-square (rms) deviation of the output operator $O_A$
actually measured from the observable $A$ to be measured,
whereas the disturbance $\eta(B)$ is defined as the root-mean-square of the change in the observable $B$
during the measurement\cite{13,14} (see {\em Methods A} for details).
The additional second and third terms result mathematically from non-commutability between $B$ ($A$) and the error (disturbance) operator (Eq.~235 in Ref.~\cite{14}).
 In particular, they imply a new accuracy limitation $\epsilon(A)\geq\frac{1}{2}\left|\bracket{\psi}{[A,B]}{\psi}\right|\sigma(B)^{-1}$
for non-disturbing ($\eta(B) = 0$) measurements and a new disturbance limitation
$\eta(B)\geq\frac{1}{2}\left|\bracket{\psi}{[A,B]}{\psi}\right|\sigma(A)^{-1}$ for noise-free ($\epsilon(A) = 0$) measurements, instead of $\epsilon(A) \sim \infty$ or $\eta(B) \sim \infty$ as derived from the Heisenberg type relation (Eq.~2).

In this letter, the universally valid error-disturbance relation (Eq.~\ref{OZA})
is experimentally tested for neutron spin-measurements\cite{Klepp08}.
We set $A$ and $B$ as the x- and y-component of the neutron $\frac{1}{2}$-spin.  (For simplicity, $\frac{\hbar}{2}$ is omitted for observables of each spin component.)
The error $\epsilon(A)$ and the disturbance $\eta(B)$ are defined for a measuring apparatus called M1, so that
the apparatus M1 measures the observable $A=\sigma_x$ with error $\epsilon(A)$ and disturbs
the observable $B=\sigma_y$ with disturbance $\eta(B)$ during the measurement.
To control the error $\epsilon(A)$ and the disturbance $\eta(B)$, the apparatus M1 is designed to actually carry out the
projective measurement of $O_A=\sigma_{\phi}=\sigma_x \cos\phi+\sigma_y\sin\phi$
instead of exactly measuring $A=\sigma_{x}$ by detuning the azimuthal angle $\phi$ of $\sigma_{\phi}$,
which is an experimentally controlled parameter, so that
$\epsilon(A)$ and $\eta(B)$ of are determined as a function of $\phi$
(see {\em Methods B} for details).
Since the output operator $O_A$ and the observable $A$ to be measured are not simultaneously measurable,
their difference is not a directly detectable quantity, and likewise the same is true for the change in the observable $B$ during
the measurement.  On this ground, the notions of the error $\epsilon(A)$ and the disturbance $\eta(B)$ have been
often claimed to be experimentally inaccessible\cite{16,17}.
In order to overcome the alleged experimental inaccessibility,
we follow the theoretical analysis (p.~387 in Ref.~\cite{14}) which proposes a method to determine the error $\epsilon(A)$ and the disturbance $\eta(B)$ from experimentally available data.
A different proposal for the experimental demonstration of the same relation (Eq.~\ref{OZA}), which exploits the weak-measurement technique, was recently published by Lund and Wiseman\cite{15+1}.

The error $\epsilon(A)$ is determined by the data from the apparatus
M1, and the disturbance $\eta(B)$ is determined by the data from
another apparatus called M2 which carries out the projective
measurement of $B$ on the state just after the M1-measurement. Thus,
the experiment is based on the successive projective measurements of
two non-commuting observables $O_A$ in M1 and $B$ in M2 as depicted
in Fig.~1.

\begin{figure}[!h]
\includegraphics[width=14.5cm]{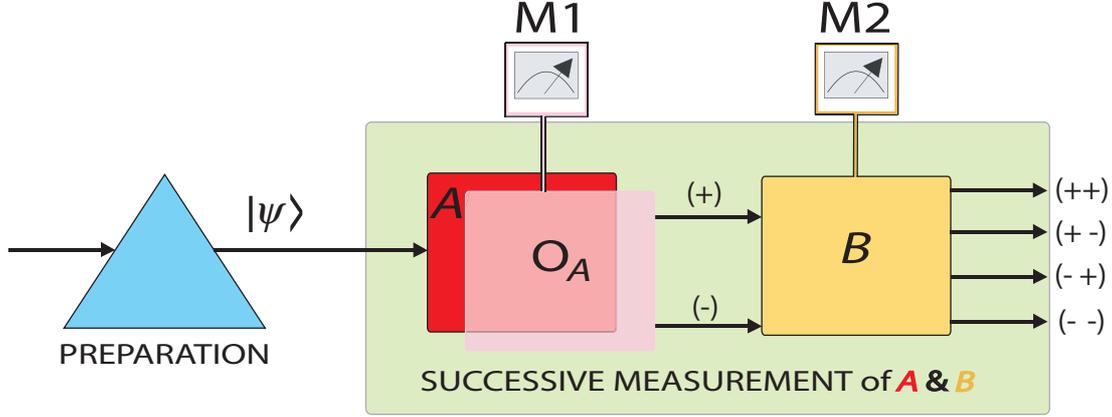}
\caption{ Experimental concept for the demonstration of the
error-disturbance uncertainty relation in successive measurements:
After preparing an initial state $\kket{\psii}$, apparatus M1 is
assumed to measure an observable $A$ (red region). The error
$\epsilon(A)$ in the $A$-measurement is experimentally controlled by
detuning apparatus M1 to measure $O_A$ instead of $A$ (light red).
After this measurement, the state is projected onto one of the
eigenstates of $O_A$, inevitabely influencing the  subsequent
measurement of observable $B$, performed by apparatus M2 (yellow
region). The disturbance $\eta(B)$ on the $B$-measurement depends on
the detuning of M1, i.e. the error of the $A$-measurement. The
successive spin-$\frac{1}{2}$ measurements of $O_A$ and $B$ result
in four possible outcomes, denoted as ($+\,+$), ($+\,-$), ($-\,+$)
and ($-\,-$), from which error $\epsilon(A)$ and disturbance
$\eta(B)$ are quantitatively determined.} \label{Scheme}
\end{figure}

For these measurements, the neutron beam first passes the preparation stage of the initial state $\kket{\psii}$.
The apparatus M1 has two possible outcomes, i.e. $+1$ and $-1$, corresponding
to measurement operators $E^{\phi}(\pm 1)=\frac{1}{2}(I\pm \sigma_{\phi})$.
The disturbance on the observable $B$ caused by the apparatus M1 is detected by the apparatus M2,
which also yields either $+1$ or $-1$, corresponding to measurement operators
$E^{y}(\pm 1)=\frac{1}{2}(I\pm \sigma_{y})$.
Thus, the successive measurements carried out by M1 and M2 finally result in four intensities denoted as ($+\,+$), ($+\,-$), ($-\,+$), and ($-\,-$).
The setup of the neutron spin experiment is depicted in Fig.~2.
The azimuthal angle $\phi$ of $\sigma_{\phi}$ is tuned between 0 and $\frac{\pi}{2}$ in a way
that a trade-off for the error and the disturbance occurs. As seen from Eq.(\ref{err}) and Eq.(\ref{dist}) error $\epsilon(A)$ and disturbance $\eta(B)$ are obtained by performing the successive measurement of $O_A$ and $B$ on different states. For measuring error and disturbance in the state $\ket{\psi}=\ket{+z}$, the auxiliary states $\ket{-z}$, $\ket{+x}$ and $\ket{+y}$ are prepared likewise (see \textit{Methods C} for more details).


\begin{figure}[!t]
\includegraphics[width=13cm]{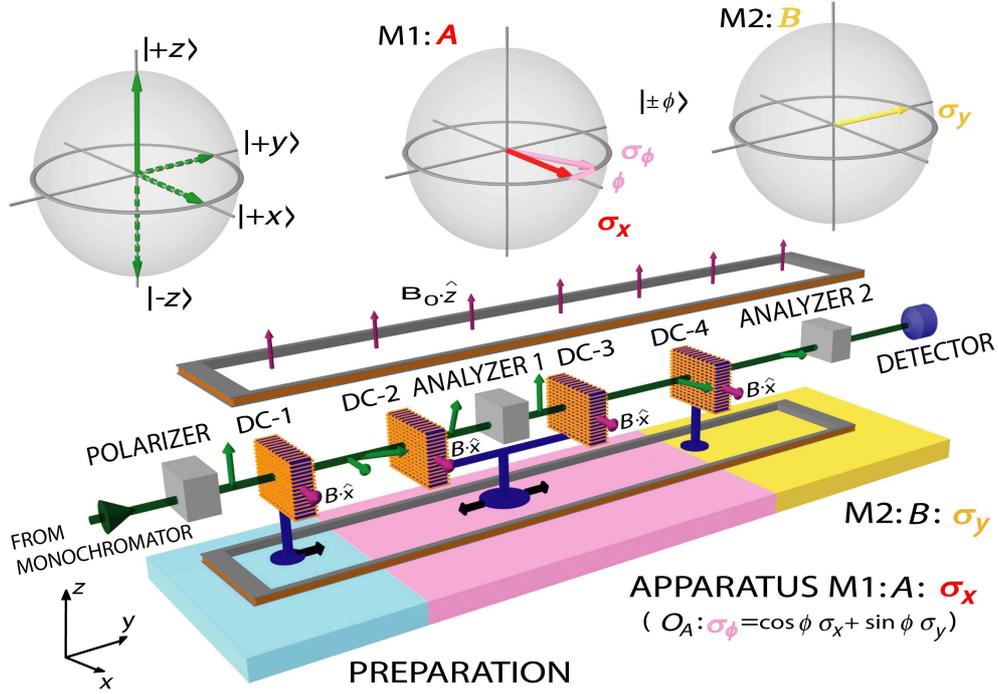}\label{fig2}
\caption{Illustration of the experimental setup for demonstration of
the universally valid uncertainty relation for error and disturbance
in neutron spin-measurements. The neutron optical setup consists of
three stages: preparation (blue region), apparatus M1 performing
measurement of observable $O_A=\sigma_\phi$ (red region) and
apparatus M2 carrying out measurement of observable $B=\sigma_y$
(yellow region). In the preparation stage a monochromatic neutron
beam is highly $(99\%)$ polarized in +z-direction by passing through
a super-mirror spin polarizer. The first DC coil (DC-1) produces a
magnetic field in x-direction ($B_x$) which can be used to rotate
the inital polarization vector about the x-axis. By additionally
exploiting Larmor precession about the +z axis, induced by the
static guide field present throughout the entire setup, and varying
the position of DC-1 arbitrary initial spin states can be produced
at end of the preparation stage (up to an irrelevant phase factor).
The projective measurement of observable $O_A$ is realized by
similar components: the prepared state rotates about the z-axis due
to Larmor precession. Hence, by properly placed DC-2 coil the spin
component to be measured can be projected towards the +z direction,
where it is reflected by a super-mirror analyzer (Analyzer-1). After
passing through this first analyzer in the +z state, DC-3 produces
the eigenstate $\ket{\pm\phi}$ of $\sigma_\phi$. In the same manner,
apparatus M2 performs the measurement of observable $B$ on the
eigenstate $\ket{\pm\phi}$, which constitutes the source of the
disturbance on the second measurement. The combination of the
projective measurements of $\sigma_{\phi}$ and $\sigma_{y}$ gives
four count rates at the neutron detector in the downstream of the
beam. The error $\epsilon(A)$ and the disturbance $\eta(B)$, as well
as the standard deviations of each measurement $\sigma(A)$ and
$\sigma(B)$, are determined from the expectation values of the
successive measurement. } \label{Setup}
\end{figure}

The experiment was carried out at the research reactor facility TRIGA Mark II of the Vienna University of Technology (TU Vienna). The monochromatic neutron beam with a mean wavelength of 1.96 $\mathring{A}$ propagates in
+y-direction. The beam is approximately $99\%$ polarized crossing a bent Co-Ti super-mirror array (polarizer)\cite{Williams}.
Two analyzing super-mirrors (analyzers) are adjusted to higher incident angles so that the second order harmonics in the incident beam are suppressed. The final intensity was about 90 neutrons/s at a beam cross section of
10 (vertical) $\times$ 5 (horizontal) mm$^2$. A $^{3}$He monitor detector is used for normalization in order to correct statistical fluctuations evoked by the reactor power. A BF$_3$ detector with high efficiency (more than $99\%$) is used for the experiment. To avoid unwanted depolarization a static guide field pointing in +z-direction with a strength of about 10 Gauss permeates rectangular Helmholtz coils. In addition, the guide field induces Larmor precession, which, together with four appropriately placed DC spin rotator coils, allows state preparation and projective measurements of $O_A$ in M1 and $B$ in M2.
(see {\em Methods D} for more details).

\begin{figure}
  \centering
  \includegraphics[width=0.79\textwidth]%
    {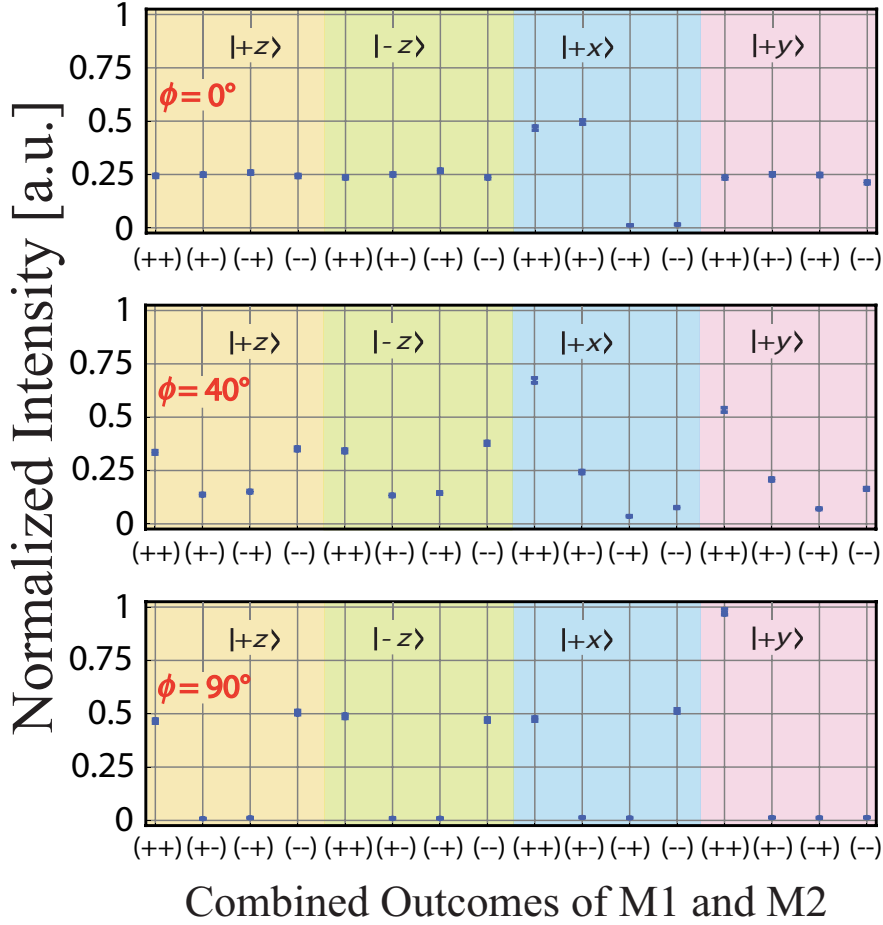}
  \caption{Normalized intensity of the successive measurements carried out by apparatus M1 and M2. The successive measurement of M1 and M2 has four outcomes, denoted as ($+\,+$), ($+\,-$), ($-\,+$) and ($-\,-$). Intensities, according to the corresponding outcomes, are depicted for each initial spin states, i.e. $\ket{+z}$, $\ket{-z}$, $\ket{+x}$ and $\ket{+y}$. Three sets for detuning parameter $\phi=0,40$ and $90$\,deg are plotted. The error $\epsilon(A)$ and the disturbance $\eta(B)$ are determined from these 16 intensities, for each setting of the detuning parameter
  $\phi$.}\label{Counts}
\end{figure}

To test the universally valid uncertainty relation stated in Eq.~\ref{OZA}, the standard deviations $\sigma(A)$, $\sigma(B)$, the error $\epsilon(A)$ and the disturbance $\eta(B)$ are determined.
%
 The measurement of the standard deviations $\sigma(A)$ and $\sigma(B)$ is carried out by M1 and M2 separately, whereas error $\epsilon(A)$ and disturbance $\eta(B)$ are determined by successive projective measurements utilizing M1 and M2. Typical experimental data sets, for miscellaneous detuning angles $\phi$, are depicted in Fig.\,\ref{Counts}.

\begin{figure}
\includegraphics[width=5cm]{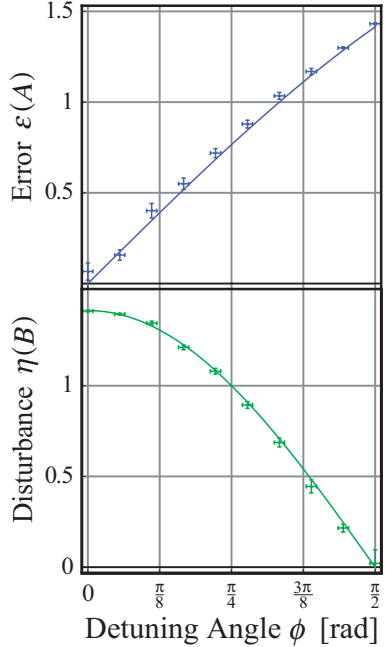}
\caption{The trade-off relation between the error $\epsilon(A)$
(blue), and the disturbance $\eta(B)$ (green), obtained as a
function of detuning the azimuthal angle $\phi$. The theory predicts
the dependence of $\phi$: $\epsilon(A)=2\sin\frac{\phi}{2}$ and
$\eta(B)=\sqrt{2}\cos\phi$. The values at $\phi=0$ are
$\epsilon(A)=0$ and $\eta(B)=\sqrt{2}$ whereas the values at
$\phi=\frac{\pi}{2}$ are $\epsilon(A)=\sqrt{2}$ and $\eta(B)=0$. It
is impossible to accurately measure both observables $A$ and $B$ at
the same time. Vertical and horizontal error bars include
statistical and systematical errors.} \label{tradeoff}
\end{figure}

 The resulting values of $\epsilon(A)$ and $\eta(B)$, together with the theoretical predictions $\epsilon(A)=2\sin\frac{\phi}{2}$ and $\eta(B)=\sqrt{2}\cos\phi$ (see {\em Methods B}), are plotted as a function of the detuning parameter $\phi$ in Fig.\,\ref{tradeoff}. The trade-off relation of $\epsilon(A)$ and $\eta(B)$ is in good agreement with theory: when one observable is measured more precisely, the other is more disturbed. The vertical and horizontal error bars in Fig.\,\ref{tradeoff} contain the statistical fluctuations of the measurement as well as the systematical misalignments of coil position and current values resulting in an angle deviation of 1.6 degrees. The final results were obtained by taking the contrast $(\sim 96\%)$ of the entire measurement into account.

\begin{figure}[!ht]
\includegraphics[width=6cm]{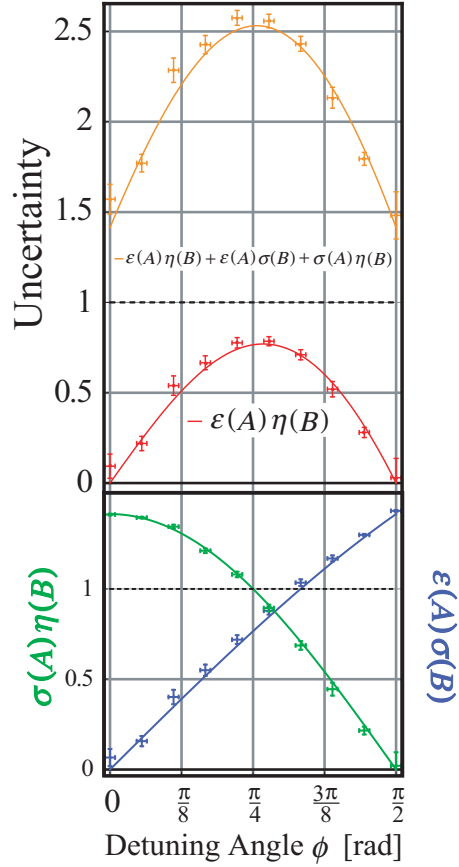}
\caption{Experimentally determined values of various products
consisting of error $\epsilon$, disturbance $\eta$, and standard
deviations $\sigma$ together with theoretical predictions. (Upper
panel) The left side of Eq.(\ref{OZA})
$\epsilon(A)\eta(B)+\epsilon(A)\sigma(B)+\sigma(A)\eta(B) =
 2\sqrt{2}\sin\frac{\phi}{2}\cos\phi
+2\sin\frac{\phi}{2} +\sqrt{2}\cos\phi$ (orange) and the Heisenberg
product $\epsilon(A)\eta(B)= 2\sqrt{2}\sin\frac{\phi}{2}\cos\phi$
(red) are plotted as a function of the detuned azimuthal angle
$\phi$. The Heisenberg product is always smaller than the calculated
limit
$\frac{1}{2}\left|\bracket{\psii}{[A,B]}{\psii}\right|=\frac{1}{2}\left|\left\langle
+z\left|\left[\sigma_x,\sigma_y\right]\right|+z\right\rangle\right|=1
$ (depicted as the dashed line). In contrast, the new sum is always
larger than the limit. This is a clear evidence of the violation of
the old uncertainty relation, in a sense of Heisenberg's
error-disturbance relation, and the validity of the new relation
consisting of the three-terms. (Lower panel) The two additional
product-terms $\sigma(A)\eta(B)$ (green) and $\epsilon(A)\sigma(B)$
(blue) in the new relation are plotted together with the theoretical
predicted curves: $\epsilon(A)\sigma(B)=2\sin\frac{\phi}{2}$ and
$\sigma(A)\eta(B)=\sqrt{2}\cos\phi$. Error bars of the experimental
data include statistical and systematical errors.} \label{fig:Ozawa}
\end{figure}

From the terms obtained above (error $\epsilon(A)$, disturbance $\eta(B)$, standard deviations $\sigma(A)$ and $\sigma(B)$), the Heisenberg error-disturbance product $\epsilon(A)\eta(B)$ and the left side of the new relation (Eq.~\ref{OZA}) $\epsilon(A)\eta(B)+\epsilon(A)\sigma(B)+\sigma(A)\eta(B)$ are plotted as a function of the detuned azimuthal angle $\phi$ in the upper panel of Fig.\,\ref{fig:Ozawa}. This figure illustrates the fact that the Heisenberg product is always below the calculated limit, and that the new sum is always larger than the limit in the scanned range of $\phi$. This clearly demonstrates that the Heisenberg-type error-disturbance relation (Eq.~\ref{HEIS}) with the single product of the error $\epsilon(A)$ and disturbance $\eta(B)$ is violated whereas the new relation (Eq.~\ref{OZA}) consisting of three terms is always satisfied.

The technique utilizing various different incident states for the estimation of the effects of the quantum operation reminds us of quantum process tomography\cite{30}. Although a parameter for the error is experimentally controlled in the experiments, it is easily extendable for uncontrolled and fluctuating parameters. Here, we concentrate on the situation where the full trade-off relation between error $\epsilon(A)$ and disturbance $\eta(B)$ occurs.
It is worth noting that the mean value of the observable $A$ is correctly reproduced for any detuning angle $\phi$,
i.e., $\left\langle +z\left|O_A\right|+z\right\rangle=\left\langle +z\left|A\right|+z\right\rangle$,
so that the projective measurement of $O_A$ reproduces the correct probability distribution of $A$,
while we can detect the non-zero rms error $\epsilon(A)$ for $\phi\ne 0$.
What has been accepted by the uncertainty principle is the existence of an unavoidable trade-off
between measurement accuracy and disturbance, but this principle has eluded a satisfactory
quantitative description for a long time. Our result is the first evidence for the validity of the new relation (Eq.~\ref{OZA}) proposed as a universally valid error-disturbance relation by one of the present authors\cite{13,14}, whereas the failure of the old relation (Eq.~\ref{HEIS}) is also illustrated.  Our results witness that the new relation solves a long-standing problem of describing the relation between measurement accuracy and disturbance.
Our demonstration sheds a light not only on fundamental limitations of quantum measurements but also on
technical limitations of precise measurements such as gravitational-wave-detection\cite{9,10,11,12} and quantum information processing.


\section*{Methods}

\subsection*{A) Universally valid uncertainty relation}

Any measuring apparatus $\bf{M}$ is, in principle, modeled by the unitary operator $U$ describing the time
evolution of the composite system of the measured object $\bf{S}$ and the probe system $\bf{P}$ during the measuring interaction and the meter observable $M$ of $\bf{P}$ actually measured after the measuring interaction\cite{14}.
If the initial states of the object and the apparatus is $\kket{\psi}$
and $\kket{\xi}$, respectively, the root-mean-square (rms) error $\epsilon(A)$ of $\bf{M}$ for measuring an observable
$A$ of $\bf{S}$ and the rms disturbance $\eta(B)$ of $\bf{M}$ caused on an observable $B$ of $\bf{S}$
is defined as
\begin{eqnarray}
\epsilon(A)&=&\|\,[U^{\dagger}(I\otimes M)U-A\otimes I]\kket{\psi}\kotimes\kket{\xi}\|,\\
\eta(B)&=&\|\,[U^{\dagger}(B\otimes I)U-B\otimes I]\kket{\psi}\kotimes\kket{\xi}\|.
\end{eqnarray}
Then, it is mathematically proved\cite{13,14} that Eq.~\ref{OZA} holds for any unitary operator $U$ for $\bf{S}+\bf{P}$,
observable $M$ of $\bf{P}$, and state vector $\kket{\psii}$ of $\bf{S}$ and
$\kket{\xi}$ of $\bf{P}$.
Suppose that the apparatus $\bf{M}$ has a family $\{M_m\}$ of the measurement operators\cite{30}.
This means that the measuring apparatus $\bf{M}$ has possible outcomes $m$ with probability $p(m)=\|M_{m}\kket{\psi}\|^2$
and the state of the object $\bf{S}$ after the measurement is $M_{m}\kket{\psii}/\|M_{m}\kket{\psii}\|$.
In this case, the rms error and rms disturbance are give by\cite{31}
\begin{eqnarray}
\epsilon(A)^2&=&\sum_{m}\|M_{m}(m-A)\kket{\psi}\|^{2},\label{e-mof}\\
\eta(B)^2&=&\sum_{m}\|[M_m,B]\kket{\psi}\|^{2}.
\label{d-mof}
\end{eqnarray}
If $\{M_m\}$ consists of orthogonal projections, the measurement is called a projective measurement.
In this case, Eq.~\ref{e-mof} can be simplified as $\epsilon(A)=\|(O_A-A)\kket{\psi}\|$ by the Pythagorean theorem,
where $O_A=\sum_{m}mM_{m}$ is called the output operator.

\subsection*{B) Error and disturbance in spin measurements: Theoretical determination}
 In the experiment,  we test the universally valid uncertainty relation (Eq.~\ref{OZA}) for observables
$A= \sigma_x$ and $B= \sigma_y$, while the initial state $\kket{\psii}$ is
$\left|+z\right\rangle$ and
the measuring apparatus ${\bf M}={\rm M1}$
is considered to carry out the projective measurement of $O_A= \sigma_\phi=\cos\phi\sigma_x+\sin\phi\sigma_y$.
Thus, the apparatus M1 is described by measurement operators
$E^{\phi}(+1)=(1+\sigma_{\phi})/2$ and $E^{\phi}(-1)=(1-\sigma_{\phi})/2$ with
$O_A=\sum_{x=\pm1}xE^{\phi}(x)$.
From Eqs.~(\ref{e-mof}) and (\ref{d-mof}), we have
\begin{eqnarray}
\epsilon(A)&=&\|(\sigma_{\phi}-\sigma_{x})\kket{\psi}\|=2\sin\frac{\phi}{2},\\
\eta(B)&=&\sqrt{2}\|[\sigma_{\phi},\sigma_{y}]\kket{\psi}\|=\sqrt{2}\cos\phi.
\end{eqnarray}

\subsection*{C)Experimental determination of error and disturbance}
In the experiment, we determine $\epsilon(A)$ and $\eta(B)$
from statistically available data obtained by successive neutron spin-measurements.
According to the former theoretical analysis (Ref.~\cite{14}, p.~387),
the error $\epsilon(A)$ is determined by mean values of $O_A$ in three
different states as
\begin{eqnarray}
    \epsilon(A)^2&=& \bracket{\psii}{A^2}{\psii}+\bracket{\psii}{O_A^2}{\psii}
    +\bracket{\psii}{O_A}{\psii}
    +\bracket{\psii}{AO_AA}{\psii}-\bracket{\psii}{(A+I)O_A(A+I)}{\psii}\nonumber \\
&=&2+\bracket{\psii}{O_A}{\psii}+\bracket{A\psii}{O_A}{ A\psii}-
\bracket{(A+I)\psii}{O_A}{(A+I)\psii},
    \label{err}
\end{eqnarray}
where we have used the following abbreviations:
$\kket{A\psii}=A\kket{\psii}$ and
$\kket{(A+I)\psii}=(A+I)\kket{\psii}$.
Since the apparatus M1 carries out the projective measurement of $O_A$,
in order to determine $\epsilon(A)$ for the basic initial state $\kket{\psii}$
we need only to measure the intensities from the apparatus
M1 in the three auxiliary incident states of M1 corresponding to $\kket{\psii}, A\kket{\psii}, (A+I)\kket{\psii}$. The expectation values expressed in Eq.(\ref{err}) are calculated from the measured intensities, depicted in Fig.\,\ref{Counts}, via
\begin{eqnarray}
\bracket{\psi}{O_A}{\psi}=\frac{(I_{++}+I_{+-})- (I_{-+}+I_{--})}{I_{++}+I_{+-}+I_{-+}+I_{--}}.
\end{eqnarray}
In order to detect the disturbance on $B$ caused by the apparatus M1, the apparatus M2 carries out
the projective measurement of $B$ in the state just after the M1-measurement.
The modified output operators of the apparatus M2 for the initial state of M1 is given by
$O_{B}=\sum_{x}E^{\phi}(x)BE^{\phi}(x)$ and
$O_{B}^{(2)}=\sum_{x}E^{\phi}(x)B^2E^{\phi}(x)$,
which describes the mean and the second moment of the observable $B$
for the initial state of M1.
Then, from Eqs.~(189) and (227) of Ref.~\cite{14}
the disturbance $\eta(B)$ is also determined by mean values of $O_B$ in three different states as
\begin{eqnarray}
\eta(B)^2&=& \bracket{\psii}{B^2}{\psii}+\bracket{\psii}{O_{B}^{(2)}}{\psii}+
\bracket{\psii}{O_{B}}{\psii}
+\bracket{B\psii}{O_{B}}{B\psii}
-\bracket{(B+I)\psii}{O_{B}}{(B+I)\psii}\nonumber\\
&=&2+\bracket{\psii}{O_{B}}{\psii}+\bracket{B\psii}{O_{B}}{B\psii}-
\bracket{(B+I)\psii}{O_B}{(B+I)\psii}, \label{dist}
\end{eqnarray}
where the illegitimate notations $\kket{B\psii}$ and $\kket{(B+I)\psii}$
are used as before and the expectation values are given by
 \begin{eqnarray}
 \bracket{\psi}{O_B}{\psi}=\frac{(I_{++}+I_{-+})- (I_{+-}+I_{--})}{I_{++}+I_{+-}+I_{-+}+I_{--}}.
\end{eqnarray}
Thus $\eta(B)$ is determined in the same manner as $\epsilon(A)$.
By the relations $\kket{\psii}=\ket{+z}$, $A\kket{\psii}=\ket{-z}$,
$B\kket{\psii}=i\ket{-z}$,  $(A+I)\kket{\psii}=\sqrt{2}\ket{+x}$,
and $(B+I)\kket{\psii}=\sqrt{2}\ket{+y}$, where we set $\ket{+x}=(\ket{+z}+\ket{-z})/\sqrt{2}$
and $\ket{+y}=(\ket{+z}+i\ket{-z})/{\sqrt{2}}$,
the required spin-states in Eqs.~\ref{err} and \ref{dist} are generated
by spinor-rotations in the experiment (see Fig.\ref{Setup}); the normalization factors are confirmed experimentally in spin-rotation measurement.

\subsection*{D) Successive neutron spin-measurements}

The spin state of the neutrons is controlled by four DC coil spin-turners. The required
incident states are prepared by first DC coil (DC-1). DC-1 is switched off for the generation of $\left|+z\right\rangle$ incident state and the spin is flipped by DC-1 for $\left|-z\right\rangle$ preparation. While the incident spin state $\left|+y\right\rangle$ is generated by applying a $\frac{\pi}{2}$ rotation around the x-axis, the incident spin state $\left|+x\right\rangle$  is produced by additionally
moving the position of DC-1 one quarter of Larmor rotation period due to the guide field. The apparatus M1 consists of
the combination of the spin-turner coils, DC-2 and DC-3, the guide field and the spin analyzer 1. The spin analyzer in our experiment performs the projective measurement of the spin's +z-component. Instead of rotating the analyzer, the neutron's spin-component in the x-y plane is unitarily rotated towards the analysis-direction by DC-2 and the guide field. DC-3 finally generates the eigenstate $\ket{\pm\phi}= E^{\phi}(\pm 1)\kket{\psii}$ (up to phase factor),
so that the apparatus M1 performs the projective measurement of $O_A$ to obtain the mean values of  $O_A$
in Eq.(\ref{err}).
The detuning of $A$ to $O_A$ is adjusted by shifting DC-2 and DC-3 likewise towards or away from the analyzer 1.
The apparatus M2 consists of DC-4 and analyzer 2 and
performs the projective measurement of $B$ on the state just after the M1-measurement
to evaluate the disturbance on $B$ caused by M1.
The coil DC-4 turns the y-component of the spin into +z-direction by $\frac{\pi}{2}$ rotation around the x-axis,
with which analyzer 2 performs the second projective measurement of $B$;
additional spin-rotation towards the y-direction is omitted here, since
only (spin-insensitive) intensity measurement is performed afterwards.
By the disturbance caused by the apparatus M1, the passage until the M2 measurement is described by
the output operator $O_B$ for the incident state of M1, and hence we obtain the mean values of $O_B$
in Eq.(\ref{dist}) from the apparatus M2.
Note that the measurement performed by apparatus M2 is the error-free $B$-measurement on the state just after the measurement carried out by M1.

\begin{acknowledgments}

We acknowledge support by the Austrian Science Fund (FWF), the European Research Council (ERC), the Japan Science and Technology Agency (JST) and The Ministry of Education, Culture, Sports, Science and Technology  (MEXT) in Japan.
 We thank H. Rauch, M. Arndt (Vienna) and A. Hosoya (Tokyo) for their helpful comments.

\textbf{Author Information}  Reprints and permissions information is available at www.nature.com/reprints. Correspondence and requests for materials should be addressed to Y.H. (Hasegawa@ati.ac.at).

\textbf{Author contributions}

J.E., G.S. and S.S. carried out the experiment and analyzed the data; G. B. contributed to the development at the early stage of the experiments; M.O. supplied the theoretical part; Y.H. conceived and performed the experiment; J.E., G.S., S.S. and Y.H. contributed the experimental set-up; and all authors co-wrote the paper.

\end{acknowledgments}


\end{document}